\documentclass[12pt,a4paper]{article}
\newcommand{\citen}{\cite}

\usepackage{graphicx}
\usepackage[]{hyperref}

\def\vx{\mathbf{x}}

\def\vL{\mathbf{L}}
\def\vr{\mathbf{r}}
\def\mvL{|\vL|}
\def\vu{\mathbf{u}}
\def\vz{\mathbf{z}}
\def\vp{\mathbf{p}}
\def\vq{\mathbf{q}}
\def\vqs{|\vq|}
\def\vk{\mathbf{k}}
\def\vQ{\mathbf{Q}}

\begin{document}

\begin{flushright}
INR-TH-2024-009
\end{flushright}

\vspace{10pt}
\begin{center}
  {\LARGE \bf
Toward a quantum field theoretical description of  oscillation effects
} \\
\vspace{20pt}
Maxim Libanov$^{a,b}$  \\
\vspace{15pt}

$^a$\textit{
Institute for Nuclear Research of
         the Russian Academy of Sciences,\\  60th October Anniversary
  Prospect, 7a, 117312 Moscow, Russia}\\
\vspace{5pt}

$^b$\textit{Moscow Institute of Physics and Technology,\\
Institutskii per., 9, 141700, Dolgoprudny, Moscow Region, Russia
}
\\
\textit{ml@ms2.inr.ac.ru}
    \end{center}
    \vspace{5pt}


\begin{abstract}
In this paper, we propose an approach based on the theory of an axiomatic $S$
matrix and partially switching on an interaction, which is extremely suitable
for describing the phenomenon of oscillations within the framework of quantum
field theory. We discuss the relation of the proposed approach with other
approaches based on  considering of wave packets as asymptotic states or
 introducing of ``distance- or time-dependent propagators''.
\end{abstract}



\section{Introduction}

One of the most interesting effects of particle physics is the phenomenon of
oscillations. This effect appears and is observed in the physics of
neutrinos~\cite{Pontecorvo:1957cp, Gribov:1968kq}, K and B
mesons~\cite{Pais:1955sm, Okun:1975di}, and should also manifest itself in the
transitions of photons into hypothetical axion-like
particles~\cite{Raffelt:1987im}. The key point for the appearance of
oscillations is the presence in the model of several freely propagating degrees
of freedom, which are the eigenstates of a free Hamiltonian with different
masses $m_{i}$. Such states form a mass basis $|\phi _{i}\rangle $. However,
due to the fact that the interaction Hamiltonian is non-diagonal in such a
basis, such states cannot be produced or registered perse. Only their linear
combinations, the so-called flavor states,
\begin{equation}
|\phi _{\alpha }\rangle =U^{*}_{\alpha i}|\phi _{i}\rangle , \ \ \ |\phi
_{i}\rangle =U_{\alpha i}|\phi _{\alpha }\rangle,
\label{Eq/Pg2/1:osc}
\end{equation}
are produced or registered. These states obviously no longer have a certain
mass and, as a result, if the produced state $\phi _{\alpha }$ has, for
instance, a defined energy, then its momentum is undefined.

On the other hand, the freely evolving state $|\phi _{i}\rangle $ at the
initial moment with a defined energy $E_{i}$ and defined momentum
$|\mathbf{p}_{i}|=\sqrt{E_{i}^{2}-m_{i}^{2}}$  at time $T$ at a distance $L$
from the source has the form\footnote{
Here and in what follows
we implicitly
mean the case of neutrino oscillations. However, our discussion and
main results easely extend to other cases of oscillations. In
particular, if the mass states are unstable, like the kaons, this can be taken
into account by the replacing in the corresponding equations (e.g., in
(\ref{Eq/Pg3/2:osc})) $m_{i}^{2}\to m_{i}^{2}-im_{i}\Gamma _{i}$ where $\Gamma
_{i}$ is the decay width. }
\[
|\phi _{j}(T,L)\rangle =\mathit{e} ^{-i(E_{j}T-|\vp_{j}|L)}|\phi _{j}(0)\rangle
\simeq \mathit{e}^{-i\frac{m_{j}^{2}}{2E}L}|\phi _{j}(0)\rangle
\]
where the latter approximate equality holds for the ultrarelativistic case
$E\gg m_{i}$, which is of interest in most cases. Accordingly, one has
\[
|\phi _{\alpha }(T,L)\rangle =\sum \limits_{j}^{}U_{\alpha j}^{*}\mathit{e}
^{-i(E_{j}T-|\vp_{j}|L)}|\phi _{j}(0)\rangle .
\]
 The amplitude of the transition of the flavor state $\alpha $ to the state
$\beta $ is given by
\begin{eqnarray}
A_{\alpha \to \beta }=\langle \phi _{\beta }|\phi _{\alpha }(T,L)\rangle =\sum
\limits_{i,j}^{} U_{\beta i}\langle \phi _{i}|
\mathit{e} ^{-i(E_{j}T-|\vp_{j}|L)}|\phi _{j}\rangle U_{\alpha
j}^{*}=\nonumber\\
=\sum \limits_{j}^{}U_{\beta j}\mathit{e} ^{-i(E_{j}T-|\vp_{j}|L)}U_{\alpha
j}^{*}\simeq
\sum \limits_{j}^{}U_{\beta j}\mathit{e}^{-i\frac{m_{j}^{2}}{2E}L}U_{\alpha j}
^{*}
\label{Eq/Pg2/2:osc}
\end{eqnarray}
what leads to well-known results for the probability of oscillatory transitions.

In the literature this approach is often called a \textit{plane-wave
approximation} and in fact exploits the apparatus of quantum mechanics rather
than quantum field theory (QFT). However, despite its simplicity, this approach
suffers from inconsistencies and leads to a number of paradoxes (see, for
example, the criticism in~\citen{Naumov:2020yyv}). In particular, in the
literal
application of this approach, energy and/or momentum are not conserved. Indeed,
if the state $|\phi _{\alpha }\rangle $ is produced as a result of the
interaction of other particles with defined energies and momenta, then its
energy and momentum will be defined (we assume that in each local act of
interaction, energy and momentum are conserved). Which, by virtue of
(\ref{Eq/Pg2/1:osc}), means the definiteness of the energy $E$ and momentum
$|\mathbf{p}|$ for all $\phi _{i}$, but this contradicts the condition of the
mass shell $E^{2}-\mathbf{p}^{2}=m_{i}^{2}$ with different masses $m_{i}$.

From a physical point of view, it is clear that in a consistent description of
the oscillation effect, it is necessary to take into account that the energy
and momentum of state $\phi _{i}$ cannot lie on a mass shell. Indeed, such
"states" never appear as a free states, but are always produced and detected as
linear combination (\ref{Eq/Pg2/1:osc}), which means that they propagate
between the acts of production and detection as virtual particles whose
momentum is off shell. Therefore, a consistent description of the oscillations
requires a QFT formalism.

The standard approach to describing processes in QFT framework is based on the
introduction of the operator of the $S$-matrix and is presented in almost any
textbook on QFT (see, for example, \S20 of~\citen{BSh}). Its essence
lies in the construction of the operator $S(t,t_{0})$ by solving the operator
equation
\begin{equation}
i\frac{\partial S(t,t_{0})}{\partial  t}=H_{\mathrm{int}}\cdot S(t,t_{0}),\ \ \
\ S(t_{0},t_{0})=1,
\label{Eq/Pg3/1:osc}
\end{equation}
which is a consequence of the Schrodinger equation written in the
\textit{interaction or Dirac representation}. The operator of the $S$-matrix is
obtained by taking the limit
\[
S=\lim_{t,t_{0}\to \pm\infty }S(t,t_{0}).
\]

The amplitude of the transition between states describing free particles, which
correspond to plane waves for particles with defined energy and momentum, in
the case of oscillations of interest to us has the form
\begin{eqnarray}
A_{\alpha \to \beta }=\langle \mbox{free particles, }\beta  |S|\mbox{free
particles, }\alpha \rangle \sim \nonumber \\
\sum \limits_{j}^{}U_{\alpha j}^{*}U_{\beta j}\int \limits_{}^{}d^{4}xd^{4}y
D_{j}(y-x)\mathit{e}^{-ipx+iqy}\sim \delta ^{4}(p-q)\sum \limits_{j}^{}U_{\alpha
j}^{*}U_{\beta j}\frac{i}{p^{2}-m_{j}^{2}}   ,
\label{Eq/Pg3/2:osc}
\end{eqnarray}
where $D_{j}(x)$ is the Feynman
propagator for a particle with a defined mass $m_{j}$, and $p$ and $q$ are the
total incoming and outcoming 4-momenta correspondingly. In what follows, for
simplicity sake, we skip all unimportant factors and consider the scalar
propagator. The main ideas and results of this paper will not depend on this,
and generalizations to the case of higher spin particles and recover skipped
factors are quite trivial.

Despite the fact that $A_{\alpha \to \beta }$ in general does not vanish
 at $\alpha \neq \beta $, it does not describe the oscillations, that is,
it does not lead to (\ref{Eq/Pg2/2:osc}). The reason for this lies, of course,
in  the plane-wave approximation for initial and final particles
(note the appearance of the exponent in the integral) and integration over the
entire space-time volume. Therefore, to describe the oscillations, it is not
enough to take into account the virtuality of particles and the standard
$S$-matrix approach. It is also necessary to somehow limit the domain of the
integration in (\ref{Eq/Pg3/2:osc}), that is, take into account that the acts
of interaction do not take place in the entire space-time, but are separated.

It seems that the most natural and consistent approach is to use wave
packets~\cite{Kayser:1980pi, Giunti:1993se} (see also the
review~\citen{Naumov:2020yyv} and the references therein), which is allowed by
the standard $S$-matrix approach. The essence of this approach is to abandon
the plane-wave approximation for initial and final particles, and use  a
compact wave packet for each real particle involved in the reaction. Then the
domain(s) of the integration in (\ref{Eq/Pg3/2:osc}) is automatically limited
to the overlap of the wave packets.

Although this approach seems to be natural and reflects the physical situation,
in our opinion it suffers from the following two disadvantages. Firstly, the
use of wave packets for each particle involved in the reaction leads to
cumbersome formulas. Secondly, and more importantly, we do not have reliable
considerations of which wave packet should be used. In fact, in the
consequential approach, the wave packet in the initial state must be "prepared"
and determined by the experimental conditions. However, usually such
"preparation" is not carried out in experiments on the study of oscillations.
Therefore, in the theoretical description based on wave packets, the Gaussian
wave packet~\cite{Kayser:1980pi, Giunti:1993se} and its relativistic
generalization~\cite{Naumov:2020yyv, Naumov:2010um} are usually used. Although
such a choice seems reasonable (recall that in a Gaussian package, the product
of the uncertainty of momentum and coordinate is minimal), it is not completely
unambiguous. For example, one can imagine an experiment to study neutrino
oscillations formed as a result of the decay of a muon in a muonic atom or in a
muonium. It is doubtful that modern experiments make it possible to distinguish
such a process from the decay of a free muon described by a Gaussian packet.
Therefore, taking into account the complexity of calculations, this approach
seems to us, although consistent, but oftentimes redundant.

Thus, it seems desirable to develop a formalism for describing the oscillations
that combines the simplicity of the plane-wave approximation and the rigor of
the QFT formalism. Such an attempt was made in the
papers\footnote{The author is grateful to Apostolos Pilaftsis and Bruno
Torres for pointing out these works.}~\citen{Ioannisian:1998ch,
Karamitros:2022nnh, Torres:2020gzm}. In fact, the main ideas of the present
note repeat and coincide with the ideas of these papers. Another approach to
the problem was suggested in a series of works~\citen{Volobuev:2017izt,
Egorov:2017qgk, Volobuev:2017rnb, Egorov:2019vqv, Volobuev:2019zan,
Egorov:2021iig}. The authors, rightly noting the need to limit the domain of
the integration in (\ref{Eq/Pg3/2:osc}) and thereby take into account the
space-time picture of the processes, proposed\cite{Volobuev:2017izt} replacing
the integral in (\ref{Eq/Pg3/2:osc})
\[
\int \limits_{}^{}d^{4}xd^{4}y
D_{i}(y-x)\mathit{e}^{-ipx+iqy}\to \int \limits_{}^{}d^{4}xd^{4}yD_{i}(y-x)
\delta
^{3}(\mathbf{p}(\mathbf{y}-\mathbf{x})-\mathbf{|p|}L)\mathit{e}^{-ipx+iqy}
\]
or (equivalently) replacing the usual Feynman
propagator in the momentum representation with a ``distance-dependent"
propagator
\begin{equation}
\frac{i}{p^{2}-m_{i}^{2}}\to D_{i}(p,L)=\int \limits_{}^{}d^{4}z\mathit{e}^{ipz}
D_{i}(z)\delta ^{3}(\mathbf{p}\vz-\mathbf{|p|}L).
\label{Eq/Pg5/1:osc}
\end{equation}
In other versions~\cite{Egorov:2017qgk} of this approach, it was proposed
to replace with the ``time-dependent'' propagator
\begin{equation}
\frac{i}{p^{2}-m_{i}^{2}}\to D_{i}(p,T)=\int
\limits_{}^{}d^{4}z\mathit{e}^{ipz} D_{i}(z)\delta
(z_{0}-T).
\label{Eq/Pg5/2:osc}
\end{equation}
Of course, such substitutions look unreasonable at first glance and are at odds
with the standard $S$-matrix formalism. But the authors appeal to
Feynman~\cite{Feynman:1949zx}, who developed the diagrammatic technique
regardless of the $S$-matrix.

The aim of the present paper is to develop a formalism that allows one to
describe the effect of oscillations within the $S$-matrix framework.
In addition, we will find out what the substitutions
(\ref{Eq/Pg5/1:osc}) and (\ref{Eq/Pg5/2:osc}) correspond to and give them
meaning.

\section{Axiomatic $S$-matrix}
\label{Section/Pg5/1:osc/1}
There is a widespread opinion in the literature that the $S$-matrix approach in
QFT is not suitable for describing the oscillation effect. It's really so, if
we keep in mind the standard approach based on eq.~(\ref{Eq/Pg3/1:osc}).
Indeed, the equation (\ref{Eq/Pg3/1:osc}) does not contain an explicit
dependence on spatial coordinates and, as a result, the spatial pattern of
interaction in the $S$-matrix constructed in this way is lost. In addition, to
obtain a relativistic $S$-matrix, it is necessary to take a limit $t,t_{0}\to
\pm\infty $ at the end of calculations. As a result, all dependence on the
space-time coordinates in the $S$-matrix constructed in this way disappears. On
the other hand, a nontrivial space-time picture is required to describe the
oscillations. Therefore, in fact, the only way to get this picture back is to
take it into account in the in- and out-states. At the same time, since the
plane-wave approach does not cope with this (due to the fact that the plane
wave is defined uniformaly throughout space-time), the only way out is to use
wave packets.

However, fortunately, there is another approach to the construction of the
$S$-matrix, known as axiomatic, without reference to the Schroedinger equation.
This approach was proposed by Stueckelberg et al and formulated and developed
by Bogolyubov at the dawn of
QFT (see \S20, \S21 of textbook~\citen{BSh} and
references therein). It is based on the physical conditions explicitly
formulated for the $S$ matrix: causality, unitarity, relativistic covariance
and the  correspondence principle.

In order to give these conditions a mathematical form, it is necessary (as in
the standard approach) to resort to the operation of ``switching on'' and
''switching off'' of the interaction and replace the
interaction Lagrangian $\mathcal{ L}\to f(x)\mathcal{ L}$ where $0\leq
f(x)\leq 1$ is a function with a compact support.
It is assumed that at asymptotically large space-time distances (at infinity)
$f(x)$ tends to zero and, as a result, the $S$-matrix becomes the unit one and
the states correspond to the free particles, that is, can be described by the
plane waves if their energies and momenta are defined. However, in the
domain(s) where $f(x)$ is nonzero $S$-matrix is a functional of $f$: $S\to
S[f(x)]$. The transition amplitudes becomes functionals as well and can be
found by the following equation
\[
A\to A[f]=\langle \mbox{free particles}|S[f]|\mbox{free particles}\rangle  .
\]
The functional $S[f]$ can be constructed by making use the explicit equations
that correspond to the physical conditions mentioned above.
The result is (up to the set of
quasilocal operators which are not relevant to this discussion)
\begin{equation}
S[f]=\sum \limits_{n=0}^{\infty }\frac{i^{n}}{n!}\int \limits_{}^{}T\left(
\prod \limits_{k=1}^{n}f(x_{k})\mathcal{ L}_{\mathrm{int}}(x_{k})dx_{k}
\right)  =T\!\exp\left(i\int \limits_{}^{}d^{4}xf(x)\mathcal{
L}_{\mathrm{int}}(x) \right)  .
\label{Eq/Pg1/1:osc}
\end{equation}
The standard $S$-matrix is restored by means of a transition to the limit in
which the region within which $f=1$ is (adiabatically) extended to entire
space-time, that is
\begin{equation}
S=S[1]=\lim_{f(x)\to 1}S[f(x)].
\label{Eq/Pg7/1:osc}
\end{equation}

The key point to the following is that the generalized $S[f]$-matrix is nothing
worse than $S[1]$. By the construction, it satisfies all required physical
conditions. Thus, in general, there is no need to take the limit
(\ref{Eq/Pg7/1:osc}). On the contrary, one can use $f(x)$ to simulate a physical
situation in the problem at hands, in particular, to localize
interactions.

Let's get back to the discussion of the oscillation phenomena.

\section{Oscillations and the axiomatic $S$-matrix approach}

Let's choose the support of $f(x)$ in the source and the detector domains and
take into account that the distance between the source and the detector
$|\vL|$ is much large than the source and detector sizes. Then we write $f(x)=
g_{s}(x)+g_{d}(x-L)$, where hereafter $L^{\mu }=(T,\vL)$ is a 4-vector, and the
functions $g_{s,d}$ are nonzero in a small enough domain in a vicinity of the
origin. In this way, the integral contributing to the second order amplitude
(\ref{Eq/Pg3/2:osc}) is replaced
\begin{equation}
\int \limits_{}^{}dx dy
D_{j}(y-x)\mathrm{e}^{-ipx+iqy}\to I_{j}=
\int \limits_{}^{}dx dy
\cdot g_{d}(y-L)g_{s}(x)D_{j}(y-x)\mathrm{e}^{-ipx+iqy}.
\label{Eq/Pg1/2:osc}
\end{equation}
By making use of the change of variables $x=u-z/2$, $y=u+z/2$ one rewrites the
integral in (\ref{Eq/Pg1/2:osc}) as follows
\begin{eqnarray}
I_{j}=\int \limits_{}^{} du\, dz\cdot g_{d}\left(u+\frac{z}{2}-L
\right) g_{s}\left(u-\frac{z}{2}   \right)
D_{j}(z)\mathrm{e}^{-iu(p-q)+i\frac{z}{2}(p+q)}. 
\label{Eqn/Pg1/1:osc}
\end{eqnarray}
Now let's look at some special cases of choosing the source and
detector functions $g_{s,d}$.

\subsection{Eternal localized source and detector}
\label{Subsec/Pg7/1:osc/Eternal localized source and detector}
Let's consider well localized static source and detector. This choice is
suitable, for example, in the case of the oscillations of the reactor
neutrinos and, in fact, was considered in Ref.~\cite{Torres:2020gzm}. If
overall sizes of the source and detector much smaller then $|\vL|$ then one can
choose $g_{s}(x)\sim g_{d}(x)\sim \delta ^{3}(\vx)$, and the integral
(\ref{Eqn/Pg1/1:osc}) becomes
\begin{equation}
 I_{j}^{(1)}\sim 2\pi \delta (p_{0}-q_{0})\tilde{D}_{j}(q,\vL),
 \label{Eq/Pg8/1:osc}
\end{equation}
where
\begin{equation}
\tilde{D}_{j}(q,\vL)= \int
\limits_{}^{}dz_{0}\mathrm{e}^{iq_{0}z_{0}-i\vq\vL}D_{j}(z_{0},\vL).
\label{Eq/Pg8/2:osc}
\end{equation}
Note that, in general, the momentum is not conserved $\vp\neq\vq$ what can be
expected since $g_{s,d}$ depend on the coordinates.

The integral (\ref{Eq/Pg8/2:osc})  can be easily calculated. To this end, one
substitutes the momentum representation for the scalar propagator
\[
D_{j}(z)=\frac{i}{(2\pi )^{4}}\int
\limits_{}^{}d^{4}k\frac{\mathrm{e}^{-ikz}}{k^{2}-m_{j}^{2}+i\varepsilon }
\]
into (\ref{Eq/Pg8/2:osc})
and performs the integration firstly
over $z_{0}$, what gives $\delta (q_{0}-k_{0})$, then over $k_{0}$ and
finally over $\vk$. As a result one gets
\begin{equation}
\tilde{D}(q,\vL)=-\frac{i}{4\pi \mvL}\mathrm{e}^{-i\vq\vL+iM_{j}\mvL},
\label{Eq/Pg8/3:osc}
\end{equation}
where
\[
M_{j}=\sqrt{q_{0}^{2}-m_{j}^{2}}+i\varepsilon.
\]

The equation (\ref{Eq/Pg8/3:osc}) leads to all known expressions for
the probability of oscillatory transitions. Indeed, in this case the amplitude
(\ref{Eq/Pg3/2:osc}) of the transitions takes the form
\[
A_{\alpha \to \beta }\sim\frac{\delta (p_{0}-q_{0})}{\mvL}\sum
\limits_{{j}}^{}U^{*}_{\alpha j}U_{\beta j}\mathrm{e}^{-i\vq\vL+iM_{j}\mvL},
\]
and, so,  the probability becomes
\begin{equation}
P_{\alpha \to \beta }\sim |A_{\alpha \to \beta }|^{2}\sim\frac{1}{\vL^{2}}\sum
\limits_{{j,k}}^{}U^{*}_{\alpha j}U_{\beta j}U_{\alpha k}U_{\beta
k}^{*}\mathrm{e}^{i(M_{j}-M_{k})\mvL}
.
\label{Eq/Pg9/1:osc}
\end{equation}
The oscillations appear from the exponent in (\ref{Eq/Pg9/1:osc})
\begin{equation}
\Delta M_{jk}=\sqrt{q_{0}^{2}-m_{j}^{2}}-\sqrt{q_{0}^{2}-m_{k}^{2}}.
\label{Eq/Pg9/2:osc}
\end{equation}
Let's consider several different limits of (\ref{Eq/Pg9/2:osc}).

\begin{itemize}
 \item
Ultrarelativistic case: $q_{0}\gg m_{j},m_{k}$. In that
case, which is the case, say, for the oscillations of neutrinos, one
finds from (\ref{Eq/Pg9/2:osc})
\begin{equation}
\Delta M_{jk}\simeq -\frac{\Delta m_{jk}^{2}}{2q_{0}},
\label{Eq/Pg9/3:osc}
\end{equation}
what leads to the well known result.


 \item
Small mass differences $\Delta m_{jk}=m_{j}-m_{k}\ll m_{j,k}$.
In the relativistic limit $q_{0}\gg m_{j},m_{k}$ one again gets the equation
(\ref{Eq/Pg9/3:osc}). So let's consider nonrelativistic case:
$M=\sqrt{q_{0}^{2}-m^{2}}<m$, but still $M\gg \Delta m_{jk}$. One has
\[
\Delta M_{jk}\simeq -\frac{m_{k}\Delta m_{jk}}{M_{k}}.
\]
Note that since $\vL$ is macroscopic, the particle is almost on a mass shell,
and $M$ is (almost) coincides with the absolute value of the 3-momenta. Thus, in
the non-relativistic limit $\mvL\simeq M_{k}/m_{k}\cdot T$, and for the
phase in (\ref{Eq/Pg9/1:osc}) one finds
\[
\Delta M_{kj}\cdot \mvL\simeq \Delta m_{jk}\cdot T.
\]
This is well known result for the kaon oscillation.

\item Large mass difference $m_{j}\gg m_{k}$ and $\Delta m_{jk}\simeq m_{j}$.
The equation (\ref{Eq/Pg9/3:osc}) is still in force with the replacement
$\Delta m_{jk}\to m_{j}$.
Thus, for the case of the
oscillations of the charged leptons, e.g., for $e\leftrightarrow\mu $, one
finds for the ultrarelativistic case
\[
q_{0}\sim m_{\mu }^{2} \mvL\sim
10^{9}\cdot\mbox{TeV}\cdot\left(\frac{\mvL}{\mbox{cm}} \right).
\]
In the non-relativistic case the phase in (\ref{Eq/Pg9/1:osc}) becomes
\[
\Delta M_{jk}\mvL\simeq -m_{j}\mvL,
\]
and, hence,
\[
\mvL\sim \frac{1}{m_{\mu }}\sim 10^{-13}\cdot\mbox{cm}.
\]
In both cases, the corresponding values are far beyond the experimental
abilities. Therefore, oscillations of charged leptons are not
observed\footnote{For a more detailed discussion of why charged leptons do not
oscillate, see, for example,~\citen{Akhmedov:2007fk}.}.
\end{itemize}

Let's now reproduce the prescription of the paper~\citen{Volobuev:2017izt}. To
this end one integrates (\ref{Eq/Pg8/2:osc}) over $\vL$ along plane which lies
at the distance $L_{0}$ in the direction $\vq$ from the source,
\begin{eqnarray}
&&\int \limits_{}^{} d^{3}L\delta (\vL\vq-L_{0}\vqs)\tilde{D}(q,\vL)=\int
\limits_{}^{}d^{3}z\delta (\vz\vq-L_{0}\vqs)\tilde{D}(q,\vz)
\nonumber\\
=
&&\!\!\int
\limits_{}^{}\!\!d^{3}z\delta (\vz\vq-L_{0}\vqs)\!\!\int
\limits_{}^{}\!\!dz_{0}\mathrm{e}^{iq_{0}z_{0}-i\vq\vz}{D}(z_{0},\vz)
=\int \limits_{}^{}\!\!d^{4}z \delta (\vz\vq-L_{0}\vqs)\mathrm{e}^{iqz}D(z).
\label{Eq/Pg2/44:osc}
\end{eqnarray}
The r.h.s. is the ``distance-dependent'' propagator introduced in
(\ref{Eq/Pg5/1:osc}). Thus, we see that the prescription (\ref{Eq/Pg5/1:osc})
corresponds to the case of the point-like source and the ``plane-like''
detector located at the distance $L_{0}$ in the direction $\vq$ from the
source.
Note, however, that although (\ref{Eq/Pg2/44:osc}) coincides with the
``distance-dependent'' propagator, the integral (\ref{Eq/Pg8/1:osc}), which
contributes to the transition amplitudes, does not.
The difference is that in the paper~\citen{Volobuev:2017izt} the amplitude is
proportional to $\delta ^{4}(p-q)$, i.e. the 4-momentum is conserved, while in
the present approach the only energy is conserved.
It is clear that this is due to the fact that the position of the
source (detector) is rigidly fixed in space: the source (detector) turns on at
the point $\vr =0$ ($\vr=\vL$), which violates the invariance with respect to
space translations, and  part of the 3-momentum goes into an external system
that holds the source (detector) at this point.
To get around this and restore the law of conservation of 3-momentum, one needs
to ``release'' the source (detector) and let it move freely. However, since we
do not want to investigate the full dynamics of the source, we can proceed as
follows. One puts the source (detector) at the point $\vr$ ($\vr+\vL$), then the
integral (\ref{Eq/Pg8/1:osc}) becomes a function of $\vr$,
\[
I_{j}^{(1)}(\vr)=\mathrm{e}^{i\vr(\vp-\vq)}I_{j}^{(1)}(0).
\]
Integrating this expression over all positions $\vr$, i.e.  actually summing up
contributions from various sources (detectors) located at various points
numbered by a vector $\vr$, we arrive to $\delta ^{3}(\vp-\vq)$.
Therefore, we see that the prescription~\cite{Volobuev:2017izt} corresponds not
only point-like source and plane-like detector,
but also requires summation of the amplitudes over all source positions.

\subsection{Distributed eternal source and detector}
Above we have taken the
source and detector functions to be proportional to $\delta^{3} (\vx)$. One may
object that in this case the $S$-matrix (\ref{Eq/Pg1/1:osc}) may be bad
defined. Apart from, we seen that the 3-momentum is not conserved (see eq.
(\ref{Eq/Pg8/1:osc})). To clarify these issues let's consider distributed
source and detector functions. For simplicity we take the following functions
\begin{equation}
g_{s}(\vx)=g_{d}(\vx)=\mathrm{e}^{-\vx^{2}a^{2}}  .
\label{Eq/Pg3/1:osco}
\end{equation}
We note that the case of completely ``switching on'' interaction corresponds
to the limit $a\to 0$ while the considered above case of $\delta $-like
source and detector is reproduced in the limit
\begin{equation}
\lim_{a\to \infty } \mathrm{e}^{-\vx^{2}a^{2}} \to \left(\frac{\sqrt{\pi
}}{a} \right)^{3}\delta^{3} (\vx)  .
\label{Eq/Pg3/2:osco}
\end{equation}

Substituting (\ref{Eq/Pg3/1:osco}) into r.h.s. of eq.
(\ref{Eqn/Pg1/1:osc}), using $\alpha $-representation for the propagator
\[
\frac{i}{k^{2}-m^{2}+i\varepsilon }=\int \limits_{0}^{\infty }d\alpha
\mathrm{e}^{i\alpha (k^{2}-m^{2}+i\varepsilon )},
\]
integrating over $u_{0},z_{0}$ and $k_{0}$ we obtain the following
integral
\begin{eqnarray}
\frac{\delta (p_{0}-q_{0})}{(2\pi )^{2}}\int \limits_{0}^{\infty }d\alpha
\int \limits_{}^{}d^{3}k\,d^{3}z\,d^{3}u\,
\exp\left(-a^{2}\left(\vu+\frac{\vz}{2}-\vL
\right)^{2}-a^{2}\left(\vu-\frac{\vz}{2}
\right)^{2}+\right.\nonumber\\
\left. i(\vp-\vq)\vu-i(\vp+\vq-2\vk)\frac{\vz}{2}+i\alpha
(M^{2}-\vk^{2}) \right)  .
\nonumber
\end{eqnarray}
Then one uses the Gaussian integral
and performs the integration over $\vu,\vz,\vk$. Finally, by making change
of the variable $2a^{2}\alpha \to \alpha  $ one arrives to
\begin{eqnarray}
(2\pi )\delta (p_{0}-q_{0})\left(\frac{\sqrt{\pi }}{\sqrt{2}a}
\right)^{3}\exp\left(-\frac{(\vp-\vq)^{2}}{8a^{2}}+
\frac{i\vL(\vp-\vq)}{2} \right) \times\nonumber\\
\int \limits_{0}^{\infty }\frac{d\alpha  }{(1+i\alpha)^{3/2}
}\frac{1}{2a^{2}}\exp\left[\frac{1}{2a^{2}}\left( \frac{Q^{2}}{1+i\alpha }+(1+i\alpha
)M^{2}-Q^{2}-M^{2} -a^{4}\vL^{2}\right) \right],
\label{Eqn/Pg4/1:osco}
\end{eqnarray}
where
\[
\vQ=\frac{\vp+\vq}{2}+ia^{2}\vL,\ \ \
Q^{2}=\vQ\cdot\vQ,\ \ \ Q=\sqrt{Q^{2}}, \ \ \
M=\sqrt{q_{0}^{2}-m^{2}}+i\varepsilon.
\]
The integral in the second line of (\ref{Eqn/Pg4/1:osco}) can be calculated.
The result is
\begin{equation}
J=\frac{1}{2 a Q}\sqrt{\frac{\pi }{2}} \mathrm{e}^{-\frac{a^4
\vL^2+(M+Q)^2}{2 a^2}}
\left(\mathrm{e}^{\frac{2 M Q}{a^2}} \left[1+
\mathrm{Erf}\left(i\frac{M-Q}{\sqrt{2} a}\right)\right]-\left[1+
\mathrm{Erf}\left(i\frac{M+Q}{\sqrt{2} a}\right)   \right]\right).
\label{Eq/Pg4/2:osco}
\end{equation}
It worth to stress that in real situations $a$ is an inverse size of the source
(detector) and, so, is always macroscopic, say, $a\leq (0.1\mathrm{m})^{-1}\sim
10^{-7}\mathrm{eV}$, while the typical momenta are large enough $|\vq|,|\vp|>
1\mathrm{eV}\gg a$. Thus, the first line in eq. (\ref{Eqn/Pg4/1:osco}) can be
always replaced by $\delta ^{4}(p-q)$. In what follows we will drop this factor
taking into account that $|\vp|=\vqs$ up to order of $a$. Let's consider the
following two limits.
\begin{itemize}
 \item
\textbf{Switching on interection. $a\to 0$.} In this limit one finds from
(\ref{Eq/Pg4/2:osco})
\[
J=\frac{i}{q^{2}-m^{2}+i\varepsilon }+\mathcal{ O}(a^{2}),
\]
that is, ordinary propagator, as it should be.

 \item
\textbf{$\delta $-like source and detector. $a\to \infty $.}
In this formal limit the momentum is not conserved: the first term in the
exponent of the first line of eq. (\ref{Eqn/Pg4/1:osco}) vanishes at any
finite momenta. Hence, we do not assume that $\vp=\vq$.  Then, the
behaviour of $J$ in this limit is
\begin{equation}
J=-i \sqrt{\frac{\pi }{2}}\frac{ e^{i
\left(\mvL M-\vL\frac{\vp+\vq}{2}\right)}}{a^3 \mvL}+\mathcal{
O}\left(\frac{1}{a^{4}} \right)  .
\label{Eq/Pg5/1:osco}
\end{equation}
To restore from this equation the obtained result (\ref{Eq/Pg8/3:osc}) one
note the difference between $\delta $-like source
(detector) which have been used in
above and distributed source (detector)
(\ref{Eq/Pg3/1:osco}) by the factor $a^{3}/\pi ^{3/2}$ (see
eq.~(\ref{Eq/Pg3/2:osco})). Taking this factor into account, and
substituting (\ref{Eq/Pg5/1:osco}) in (\ref{Eqn/Pg4/1:osco}) one comes to
eq.~(\ref{Eq/Pg8/1:osc}) with $\tilde{D}(q,\vL)$  given
by eq.~(\ref{Eq/Pg8/3:osc}).
\end{itemize}

\subsection{Instant unlocalized source and detector}
\label{Subsec/Pg13/1:osc/Instant source and detector}
Let's consider the opposite case: the source and detector are unlocalized in
space, but localized in time. If the time interval, during which the source and
detector  operate, is small enough compared to $T$, then
the approximation $g_{s}(x)\sim g_{d}(x)\sim \delta  (t)$ can be used, and the
integral (\ref{Eqn/Pg1/1:osc}) becomes
\[
I_{j}^{(2)} \sim (2\pi )^{3}\delta  ^{3}(\vp-\vq)\hat{D}_{j}(q,T),
\]
where
\begin{equation}
\hat{D}_{j}(q,T)=
\int \limits_{}^{}\!\!d^{3}z\mathrm{e}^{iq_{0}T-i\vq\vz}D_{j}(T,\vz)=
\int \limits_{}^{}\!\!dz\mathrm{e}^{iqz}D_{j}(z)\delta (z_{0}-T)=
\frac{\mathrm{e}^{iT(q_{0}-\sqrt{\vq^{2}+m^{2}_{i}})}}{2\sqrt{\vq^{2}+m^{2}_{i}}}
.
\label{Eq/Pg13/1:osc}
\end{equation}
This is nothing but a ``time-dependent'' propagator (\ref{Eq/Pg5/2:osc}) from
paper~\citen{Egorov:2017qgk}.
Thus, we see that the prescription
(\ref{Eq/Pg5/2:osc}) corresponds to the case of the instantaneous switching
on/off source and detector unlocalized in the space.
In addition, to restore the conservation law of the 4-momentum, it is necessary
to perform integration of the amplitudes over all moments of switching on the
source (detector) in the same way as it was done in the section
\ref{Subsec/Pg7/1:osc/Eternal localized source and detector}.

\subsection{Other cases of choosing the source and detector functions}
\label{Subsec/Pg13/1:osc/Other cases of choosing the source and detector
functions}

For completeness, let's list a few more different cases of choosing the
source and detector functions:
\begin{itemize}
 \item
$g_{s}\sim \delta ^{4}(x),\ g_{d}\sim \delta ^{3}(\vx)$
\[
I_{j}^{{(3)}}\sim \tilde{D}_{j}(q,\vL),\,\ \ \ \
I_{j}^{{ (1)}}=2\pi \delta (p_{0}-q_{0}) I_{j}^{(3)}
\]
 \item
$g_{s}\sim \delta ^{3}(\vx),\ g_{d}\sim \delta ^{4}(x)$
\[
I_{j}^{{(4)}}\sim
\mathrm{e}^{iT(p_{0}-q_{0})}\tilde{D}_{j}((p_{0},\vq),\vL),\,\ \ \ \ I_{j}^{{
(1)}}=\int \limits_{}^{}dTI_{j}^{(4)}
\]
\item
$g_{s}\sim \delta ^{4}(x),\ g_{d}\sim \delta ^{4}(x)$
\begin{equation}
I_{j}^{{(5)}}\sim \mathrm{e}^{iqL}D_{j}(L),\,\
\ \ \ I_{j}^{{ (1)}}=2\pi \delta (p_{0}-q_{0})\int \limits_{}^{}dTI_{j}^{(5)}
\label{Eq/Pg14/1:osc}
\end{equation}
\end{itemize}
In all considered cases  the corresponding integrals
$I_{j}^{(1)},\ldots, I_{j}^{(5)}$ lead to an oscillating pattern for the
probability of the flavor transitions. However, the ocsillating exponent may be
different. In particular, in the case (\ref{Eq/Pg14/1:osc}) the phase, which
gives rise the oscillations,  is
\[
\Delta m_{jk}\cdot \sqrt{T^{2}-\vL^{2}}=\Delta m_{jk}L.
\]
Herewith the oscillations will appear if $\max(m_{j},m_{k})\cdot L\sim 1$, that
is, if the detector is deep enough in the lightcone of the source.

Also note that not every choice of a localized source and detector leads to
the oscillation. The source and detector must be well separated in space, or
in time, or both in space and time. For example, if the source localized in
time $g_{s}\sim\delta (t)$, but the detector is localized in space
$g_{d}\sim\delta ^{3}(\vx)$, the integral (\ref{Eqn/Pg1/1:osc}) takes the form
\[
I_{j}=\frac{i}{q_{0}^{2}-\vp^{2}-m_{j}^{2}}\mathrm{e}^{i\vL(\vp-\vq)},
\]
and no oscillations will appear.

Yet, in our opinion the choice of a time-localized  detector is not
physically interesting. In fact, this choice just means that the detector does
not operate almost all the time, except for a short period around of $T$, and
no one can observe anything. In this respect the ``time-dependent'' propagator
(\ref{Eq/Pg5/2:osc}), (\ref{Eq/Pg13/1:osc}) corresponds to not very interesting
physical situation.

\section{Discussion and Conclusion}

In this paper, we proposed an approach to describing the oscillation phenomenon
based on the axiomatic construction of the $S$-matrix. The axiomatic $S$-matrix,
being a functional of the extent of switching on the interaction,  by
construction satisfies such important physical principles as causality,
unitarity and relativistic covariance. Therefore, it allows us to consistently
describe the processes of particles transitions with partially localized
interaction, using free particles with a certain momentum, i.e. plane waves, as
asymptotic states.
On the other hand, to describe oscillations within the framework of quantum
field theory, a non-trivial space-time picture with well-separated interaction
domains is required. So, the axiomatic $S[f]$-matrix is extremely suitable for
describing the oscillation effect.
Another way to obtain such
a picture is to use localized wave packets for asymptotic states.
However, the latter approach, although consistent, requires rather cumbersome
calculations, and also has the disadvantage of ambiguity in choosing the
package shape. Although there is a similar drawback in the proposed approach:
the choice of the function of switching on the interaction $ f(x) $ is
ambiguous, but it is simpler computationally. Therefore, in our opinion, both
wave packets and the switching on the interaction should be considered rather
as some regularization procedures, on the details of which the observed
physical effects of oscillations should not depend, at least until the choice
of a specific wave packet is unambiguously fixed by the experimental
conditions.

It is worth noting that, as noted in~\citen{BSh}, the function $ f(x) $ can be
introduced in a more ``physical'' way, without violating the requirements  for
the $S$-matrix formulated above. For example, it can be considered as a given
external  classical field or as a given external particle flux:
a similar approach was used, e.g., in~\citen{Kobzarev:1980nk} without referring
to the axiomatic $S$-matrix.
In particular, the profile of a wave packet or a product of wave packets can be
taken as the function $f(x)$. Then, at least formally, the approach proposed here
will reproduce the results obtained within the framework of the usual
$S$-matrix approach with the using wave packets.
Despite this remark, it is nevertheless interesting to study within the
framework of the proposed approach other effects related to oscillations, for
example, decoherence.  We postpone these questions for further research.

Finally, we have clarified the meaning of the prescriptions proposed
in~\citen{Volobuev:2017izt, Egorov:2017qgk, Volobuev:2017rnb, Egorov:2019vqv,
Volobuev:2019zan, Egorov:2021iig}, and found that they correspond to not very
interesting physical situations.

\section*{Acknowledgments}

At the end of 2017, Valery Rubakov, being the editor of the IJMPA, asked me to
look at article~\citen{Volobuev:2017izt}, which already had several negative
reviews from scientists who, apparently, deal more with the phenomenological
aspects of the oscillation phenomenon. At least, that's what the author
of~\citen{Volobuev:2017izt} claimed and asked Valery to send the paper to a
referee who was more able to evaluate the theoretical aspects of his proposed
approach. I looked at the paper and realized that I had no arguments why it
should be rejected: within the framework of fairly clearly formulated
assumptions, first of all prescriptions (\ref{Eq/Pg5/1:osc}), the work was done
flawlessly. So, I gave a recommendation to publish the article ``as
it is'', and leave the controversial prescription (\ref{Eq/Pg5/1:osc}) to the
scientific community. In addition, I myself became interested in what
prescription (\ref{Eq/Pg5/1:osc}) corresponds to in the framework of the usual
$S$-matrix approach. The result of this research is the present article. Thus, I
am grateful to Valery for bringing this issue to my attention, as well as for
numerous and fruitful discussions. I am also grateful to I.~Volobuev,
V.~Egorov, D.~Naumov and V.~Naumov for useful discussions and remarks. I thank
S.~Demidov, A.~Kataev, M.~Smolyakov and S.~Troitsky for their interest in the
work.

\end{document}